\documentclass[review,authoryear,3p]{elsarticle}

\usepackage[english]{babel}
\usepackage[utf8x]{inputenc}
\usepackage[T1]{fontenc}

\usepackage{amsmath}
\usepackage{amssymb}
\usepackage{graphicx}
\usepackage[colorlinks=true, allcolors=blue]{hyperref}
\usepackage{xcolor}
\usepackage{caption}
\usepackage{subfig}
\usepackage{tabularx}
\usepackage{multirow}
\usepackage{url}
\usepackage{array}
\usepackage[toc,page]{appendix}
\usepackage[nameinlink,noabbrev]{cleveref} 
\crefname{section}{Section}{Sections}
\crefname{subsection}{Subsection}{Subsections}
\crefname{figure}{Figure}{Figures}
\crefname{table}{Table}{Tables}
\crefname{equation}{Eq.}{Eqs.}

\journal{Renewable Energy}

\newcolumntype{C}[1]{>{\centering\arraybackslash}p{#1}}

\begin{document}

\begin{frontmatter}

\title{Intra-day solar irradiation forecast using RLS filters and satellite images}

\author[LES-MVD,IIE-FING]{Franco Marchesoni-Acland}
\author[LES-MVD]{Rodrigo Alonso-Suárez\corref{cor1}}

\address[LES-MVD]{Laboratorio de Energía Solar, Facultad de Ingeniería, UDELAR, J. H. y Reissig 565, Montevideo, Uruguay}

\address[IIE-FING]{Instituto de Ingeniería Eléctrica, Facultad de Ingeniería, UDELAR, J. H. y Reissig 565, Montevideo, Uruguay}

\cortext[cor1]{Corresp. author: R. Alonso-Suarez, r.alonso.suarez@gmail.com}

\begin{abstract}
Satellite-based solar irradiation forecasting is useful for short-term intra-day time horizons, outperforming numerical weather predictions up to 3-4 hours ahead. The main techniques for solar satellite forecast are based on sophisticated cloud motion estimates from geostationary satellite images. This work explores the use of satellite information in a simpler way, namely spatial averages that require almost no preprocessing. Adaptive auto-regressive models are used to assess the impact of this information on the forecasting performance. A complete analysis regarding model selection, the satellite averaging window size and the inclusion of satellite past measurements is provided. It is shown that: (i) satellite spatial averages are useful inputs and the averaging window size is an important parameter, (ii) satellite lags are of limited utility and spatial averages are more useful than weighted time averages, and (iii) there is no value in fine-tuning the orders of auto-regressive models for each time horizon, as the same performance can be obtained by using a fixed well-selected order. These ideas are tested for a region with intermediate solar variability, and the models succeed to outperform a proposed optimal smart persistence, used here as an exigent performance benchmark.
\end{abstract}


\end{frontmatter}

\section{Introduction}\label{sec:intro}

Solar Photovoltaics (PV) has become the world's fastest growing energy technology \citep{REN21-2019}. However, achieving a high penetration of solar PV into electricity grids is a challenging task due to solar irradiance intermittency, caused by cloud dynamics. This variability affects the demand-supply balance that is required for grid operation, implying stability risks and higher management costs, and also affects the operation of electricity markets, adding uncertainty in energy transactions. Resource forecasting is one of the actions to be taken in order to mitigate the negative effects produced by solar variability. Forecasting ability enables better decision-making both in grid and markets operations, providing valuable information for cost-effective spinning reserves management, unit commitment and for establishing more accurate energy prices and quantities for trade.

The research community in solar forecasting has been growing in the last years. As a reference, Google Scholar searches reveal an upward trend with $3.7$k, $6.4$k, $13.8$k, and $16.8$k results in each of the last four quinquenniums. There are now well-established methods for operational solar irradiance forecasting \citep{DIAGNE201365}, namely Numerical Weather Prediction (NWP) models, satellite derived cloud motion estimates, and statistical (learning) procedures over time series data. NWP models are heavily used for day ahead forecasting \citep{lorenz2009irradiance, Mathiesen-Kleissl-2013, Lara-Fanego-2012, Perez-2013c, verbois2018solar}, and they have expensive computational requirements. The prediction comes as a product of the sophisticated underlying physical models. Other approaches can not use the information as effectively in such large time horizons, given that the correlations weaken with time. A comparison of NWP models' performance can be found at \cite{lorenz2009benchmarking} and \cite{perez2011evaluation}. Alternatively, statistical or machine learning methods have been mostly used over ground data \citep{Pedro-Coimbra-2012, Lauret-2015, yagli2019automatic}. High quality GHI ground measurements are not only used as a basis for the forecasts, but more importantly, they are the ground truth used on the performance evaluation phase. Some of the proposals also integrate additional exogenous variables, as noted by \cite{Voyant-2017}. When they do not rely on ground measurements, they usually appear as natural methods to combine different models \citep{lorenz2012short, Mazorra-Aguiar-2016}. Finally, satellite-based models are dominated by Cloud Motion Fields (CMF) estimations, being the work of \cite{Lorenz-2004} a classical reference. These models are intended to project clouds by means of an estimated velocity field. There are other approaches based on satellite data, involving either different ways to estimate the CMF \citep{hammer1999short, Peng-2013} or computing correlations \citep{Dambreville-2014}. \cite{perez2013solar} and \cite{Kuhnert-2013} have reported that a satellite based method outperforms various NWP models when making forecasts up to $4$ hours ahead.

As the nature of the methods is fundamentally different, they are not expected to be totally redundant with each other. A method that integrates two or more approaches will likely perform better. For example, while GHI measurements are taken at one specific point, satellite data provides information about the cloudiness on the surrounding areas, that can be exploited by forecasting techniques to reduce the prediction uncertainty. A few works exploring the combinations of the methods were conducted by \cite{aguiar2016combining, harty2019intra, marquez2013hybrid, bright2018improved, aguiar2015use}. The input information given to these methods is diverse. Some input data require preprocessing and others do not; for instance, \cite{marquez2013hybrid} use a segmented satellite image taking the cloudiness information in the form of a ladder oriented by the main cloud displacement between images. Some works use two different methodologies for the same conceptual input, e.g. using NWP outputs from a GFS driven WRF \citep{Skamarock-2008} or WRF-Solar \citep{Jimenez-2016}, or using CMF information by means of \cite{Lorenz-2004} technique or regular optical flow techniques \citep{Horn-Schunck-1981, Lucas-Kanade-1981}. Given the high uncertainties still observed in solar forecast techniques, there is a need to further analyze the combination of the various input data.


In this work we explore the combination of input sources by means of a statistical signal processing approach. In particular, we aim to combine ground measurements with satellite information, providing a detailed assessment of the combination. The time series analysis literature is vast, including Artificial Neural Networks (ANN), classical machine learning approaches (SVM, Trees), and statistical time series models. A recent comprehensive review on these methodologies can be found at \cite{Voyant-2017}. Most of these methods are suited to include scalar exogenous variables. Here, we make use of Auto-Regressive Moving-Average (ARMA) models, that have proved to work well in this problem \citep{reikard2009predicting}. More specifically, the ARMA model is formulated as an adaptive filter through the Recursive Least Squares (RLS) algorithm as in \cite{david2016probabilistic, marchesoniacland}. In \cite{NN4} and \cite{Dambreville-2014} satellite data is integrated as input to statistical models. Both approaches avoid using complex CMF methodologies and use correlations in order to decide which pixels (or block of pixels) to include. We analyze the use of a simpler satellite input with an approach that involves almost no preprocessing: taking the mean of a window of the satellite albedo image. Satellite data carries valuable information of the surroundings of the forecast site, therefore introducing present and past satellite cloudiness data is a good way of improving short-term intra-day forecasts. Combining these present and past values can be thought as weighted time-averaging. In order to compare time-averaging with spatial-averaging the size of the spatial averaging window is modified as well. It is expected that satellite information is, to some degree, redundant with solar irradiance measurements, as irradiance estimates can be inferred from satellite images \citep{Perez-2002, Rigollier-2004, Alonso-Suarez-2012, Qu-2017}. This fact is analyzed by comparing results obtained by using more observations of ground data than of satellite data and \textit{viceversa}. The procedure to select the best model, comprising the ARMA model selection, the number of lags on the satellite inputs, and the averaging window size, is presented as well. We evaluate the benefits of using different parameters (ARMA coefficients, satellite lags and window size) and compare the models against a challenging benchmark that arises from an optimal selection of smart persistences. This article demonstrates that this simple proposal works for intermediate solar variability sites, such as the region under study in this work.

The main contributions of this article can be summarized as follows:

\begin{itemize}
\item It proposes and evaluates a methodology to easily include raw satellite data (albedo) into a time series forecasting model. As shown in \cite{aguiar2015use, marquez2013hybrid} adding solar satellite estimates improves the forecasting performance. Here, a previous step is addressed, including raw satellite albedo as input, without the postprocessing added by a solar satellite model which may add uncertainty to the problem. To the best of our knowledge, the use of raw satellite information as input of solar forecasting methods has not been tested in the literature. A detailed evaluation is made, using a challenging performance upper limit (an optimal smart persistence) for the Forecasting Skill (FS) calculation.
\item It provides an assessment of the forecasting gain by adding raw satellite information to a baseline ARMA-RLS model that only uses ground measurements. A performance analysis when varying the final model's parameters is provided, in particular, the $p$ and $q$ ARMA-RLS parameters, the satellite averaging window size and the satellite past samples (satellite lags).
\item It shows that when using only ground measurements as input there is not much to be gained by fine-tuning the ARMA-RLS model's parameters. The best performance, which is achieved by setting the optimal parameters for each lead time, presents a negligible difference with the performance that can be obtained by using a few fixed auto-regressive and moving average terms for all lead times.
\item It shows that when adding satellite albedo, the utility of the ground measurements past samples as input is restricted only to the very short-term forecast horizons (up to 30 minutes ahead). Above this limit, the performance of models that use satellite albedo is insensitive to ground measurements lags. In all cases, models including satellite information, whether they include ground measurements past values or not, achieve the best performance for all time horizons.
\item It defines and uses a natural challenging persistence benchmark that is obtained from the utilization of the optimal smart persistence procedure at each lead time. This defines the best performance curve that the simple smart persistence procedure can obtain. As explained in \Cref{susec:smart_persistence}, some authors differ and use a few different benchmark definitions of persistence or smart persistence. These definitions and some relevant work in this topic are discussed in \Cref{susec:smart_persistence}, ending with the introduction of the optimal smart persistence benchmark.
\end{itemize}

This article is organized as follows: in \Cref{sec:data} the data is presented along with a description of the equipment and stations' characteristics, the exogenous variable being used and the data quality procedure. In \Cref{sec:algorithms} the RLS algorithm is introduced, with a brief mention of the advantages of the approach. In \Cref{sec:performance} the evaluation framework is presented, describing the performance metrics to be used (\Cref{susec:metrics}) and the optimal smart persistence (\Cref{susec:smart_persistence}). \Cref{sec:results} provides the results obtained with the different models and the performance analysis. The ARMA-RLS model selection is discussed in \Cref{susec:RLS_measurements} while the inclusion of satellite albedo is addressed in \Cref{susec:RLS_cloudiness}. Finally, our conclusions are summarized in \Cref{sec:conclusions}.

\section{Data}\label{sec:data}

This section describes the two types of data used in this work: global horizontal irradiance ground measurements (GHI, $G_{h}$), and Earth albedo ($\rho_{p}$) derived from visible channel GOES-East satellite images.

\subsection{Solar irradiance data}

Solar irradiance measurements recorded at seven ground stations in the south-east part of South America are used in this work. Two of these sites, the Solar Energy Laboratory (LES, \href{http://les.edu.uy/}{http://les.edu.uy/}) experimental facility at the North of Uruguay (LE) and the S\~{a}o Martinho da Serra station from the SONDA network (\href{http://sonda.ccst.inpe.br/}{http://sonda.ccst.inpe.br/}) at the South of Brasil (MS), record GHI measurements with equipment and procedures that comply with BSRN (Baseline Solar Radiation Network, \href{https://bsrn.awi.de/}{https://bsrn.awi.de/}) requirements \citep{BSRN_OpMan}, being the latter formally a BSRN site. In these sites, the GHI is measured using spectrally flat Class A pyranometers (according to the ISO 9060:2018 standard) and routine maintenance is performed on a daily basis, such as dome cleaning. The other five stations are part of the LES solar irradiance measurement network and are located on field in semi rural environments. They are equipped with spectrally flat Class A or B pyranometers for the GHI measurement and maintenance is done on a monthly basis by personal at the stations. Based on our experience, equipments' quality, calibration schemes, and maintenance schedules, we assign a global (P95) uncertainty for GHI measurements of 3\% of the average at the LE and MS sites and of 5\% in the rest. These uncertainties are way lower than the uncertainty of the forecast being evaluated in this work.

\Cref{tab:sites} presents the sites' location, data span, and some relevant measurements' characteristics, namely the GHI average value, $\overline{G}_{h}$, and the 10-minutes nominal variability, $\sigma$. The GHI average is the value that will be used to express the performance metrics as a percentage. The nominal variability is defined by \cite{Perez-2015} as the standard deviation of the changes in the clear-sky index time-series, $\sigma = \text{Std} \{ \Delta k_{c} (t)\}$. The clear-sky index is defined as,

\begin{equation}
k_{c}(t) = \frac{G_{h}(t)}{G_{h}^{\text{csk}}(t)},
\label{eq:kc}	
\end{equation}

\noindent
where $G^{\text{csk}}_{h}$ is the output of a clear-sky model. Here, the McClear model is used \citep{Lefevre-2013}, publicly available at the CAMS platform (\href{http://www.soda-pro.com}{http://www.soda-pro.com}), to calculate the clear-sky index from the GHI time series. The values provided in \Cref{tab:sites} were calculated over the 10-minutes quality-checked daylight solar irradiance data set, as explained in \Cref{susec:quality}.

These stations are representative of the subtropical temperate climate of the south-east part of South America known as Pampa Húmeda, which is classified under the updated K\"{o}ppen-Geiger climate map as Cfa \citep{Peel-2007}. This is a warm, temperate and humid climate, with hot summers. The solar variability of the region is intermediate, both in terms of inter-annual variability \citep{Alonso-Suarez-2017} and short-term variability. The latter, more relevant for this work, is quantified by nominal variability and has an average of $\sigma = 0.148$ in the region (see \Cref{tab:sites}). Hence, the results provided in this work are applicable to sites with similar climate conditions (intermediate variability and Cfa or Cfb), as Central and South-East US, non-Mediterranean Europe and East Australia, among others. For other climates or different solar variability sites, as low-variability desert sites or high-variability insular locations, results may not be extrapolable and further investigation is required. 

\begin{table*}[!ht]
\centering
\begin{small}
\def\arraystretch{1.2}
\begin{tabular}{lccccccc}
\hline
\textbf{station}&\textbf{station}&\textbf{period}&\textbf{lat.}&\textbf{lon.}&\textbf{alt.}&$\overline{\mathbf{G}}_{\mathbf{h}}$&$\mathbf{\sigma}$\\
\textbf{name}&\textbf{code}&\textbf{of time}&(deg)&(deg)&(m)&(W/m$^2$)&(--)\\
\hline
LES facility&      LE& 01/2016 -- 12/2017& -31.28& -57.92&  56& 461& 0.139\\
S\~{a}o Martinho&  MS& 01/2012 -- 12/2015& -29.44& -53.82& 489& 451& 0.149\\
Artigas&           AR& 01/2015 -- 12/2017& -30.40& -56.51& 136& 451& 0.147\\
Las Brujas&        LB& 01/2015 -- 12/2017& -34.67& -56.34&  38& 440& 0.152\\
Tacuarembó&        TA& 01/2016 -- 12/2017& -31.71& -55.83& 142& 443& 0.147\\
Rocha&             RO& 01/2016 -- 12/2017& -34.49& -54.31&  20& 425& 0.159\\
La Estanzuela&     ZU& 01/2016 -- 12/2017& -34.34& -57.69&  70& 442& 0.144\\
\hline
&&&\multicolumn{3}{c}{\textbf{all sites average}}& \textbf{445}& \textbf{0.148}\\
\cline{4-8}
\end{tabular}
\end{small}
\caption{Solar irradiance measuring sites: location, characteristics and data span.}
\label{tab:sites}
\end{table*}

The $k_{c}$ time series, at 10-minutes granularity, is the ground measurements input considered for the forecast algorithm. This is common practice in the solar forecasting field, as the GHI time series has a daily and seasonal geometrical behavior that introduces a deterministic complexity on the statistical learning approaches. This deterministic behavior can be easily eliminated by using clear-sky estimations (or even top of the atmosphere irradiance calculations), isolating the higher-rate fluctuations due to cloudiness. With this methodology, the forecasting models can be dedicated to predict the non-deterministic component of solar irradiance due to clouds dynamics, leaving the geometric part to be represented by the clear sky model.

\subsection{Satellite images}\label{susec:satellite}

GOES-East satellite visible channel images are used here by means of the Earth Albedo ($\rho_{p}$). We are not using, for instance, solar satellite estimates. The images are preferred in a non-processed version, as a way to exclude the uncertainty associated with the conversion of the Earth Albedo (mainly, cloudiness information) to solar irradiance. The satellite images used in this work were generated by the GOES12 and GOES13 satellites, which operated in the GOES-East position during the considered time period (see \Cref{tab:sites}). During that time, the GOES-East series provided irregular acquisition for South America, usually available at a rate of two images per hour. The 10-minutes time resolution for the satellite albedo was obtained via a linear interpolation of the satellite time series. Satellite gaps of more than two consecutive hours were not interpolated and were removed from the data set. Our local GOES-East satellite database has a spatial coverage of the Pampa Húmeda region and surroundings areas.

The former GOES-East satellites (GOES12 and GOES13) had a nominal spatial resolution of 1~km on their visible channel. The location of these satellites (geostationary orbit, $75$°W) results in a pixel size of about 1-2~km over the region. To include satellite information in a simple way into the forecast algorithm, an average value is calculated in a cell centered at each site. As we are interested in analyzing the forecast performance and features for different satellite spatial average sizes, different cell sizes are tested. For easy communication, we choose three different cell sizes: small, medium and large, representing each a $1~\text{min}\times1~\text{min}$, $10~\text{min}\times10~\text{min}$ and $20~\text{min}\times20~\text{min}$ latitude-longitude cells. This approximately corresponds to cell sizes of $1.9~\text{km}\times1.6~\text{km}$, $19~\text{km}\times16~\text{km}$ and $37~\text{km}\times31~\text{km}$, respectively, over the target region.

\subsection{Data filtering}\label{susec:quality}

The data quality check and filtering is as follows. First, we exclude data with solar altitude lower than 10º to avoid using early morning or late afternoon observations which present higher relative deviations due to cosine error in the measurements. This is a standard filtering procedure that ignores only $\simeq1$\% of the annual total solar energy \citep{david2016probabilistic}. Then, we remove erroneous or missing data in the measurements or the satellite time series ($3$\% of the data). Our GHI data is flagged so that a 10-minutes observation is only calculated from the 1-minute time-series if at least 7 minutes of data are available (more than $66$\% of the interval). Next, two filters associated with irradiance upper limits are applied over the GHI data set (and remove between $0.1$-$0.2$\% of the data): (i) the BSRN quality procedure to detect physically impossible and extremely rare GHI measurements \citep{BSRN-1998, Yang-2018} and (ii) the exclusion of observations with clear sky index exceeding the value 1.35. Finally, a last check is done over the variability metric ($\sigma$), discarding a few variability outliers that arise due to the previous filtering stages ($\simeq0.1$\% of the data). The filtering procedure is summarized in \Cref{tab:data} for each station and the complete data set. The last column shows the fraction of data that is being filtered. As can be seen, around $4.4$\% of the initial data is discarded by these procedures. 

\begin{table*}[ht!]
\renewcommand{\arraystretch}{1.2}
\centering
\caption{Quality check and data set description for each measurements station.}
\begin{small}
\begin{tabular*}{\textwidth}{c@{\extracolsep{\fill}}cccccccc}
\hline
&\textbf{solar alt.}~$\mathbf{>10}$\textbf{º}&
\multicolumn{2}{c}{\textbf{missing or erroneous}}&
\multicolumn{2}{c}{\textbf{upper limit filters}}&
\multicolumn{2}{c}{\textbf{variability filter}}&
\textbf{filtered}\\
\cline{3-4}\cline{5-6}\cline{7-8}
\textbf{site}&\textbf{samples}&
samples&(\%)&samples&(\%)&samples&(\%)&
\textbf{(\%)}\\
\hline
LE& 46198& 44585& 3.5\%& 44530& 0.1\%& 44318&    0.5\%& 4.1\%\\
SM& 92541& 85528& 7.6\%& 85377& 0.2\%& 85120&    0.3\%& 8.1\%\\
AR& 69335& 66941& 3.5\%& 66836& 0.2\%& 66820& < 0.05\%& 3.6\%\\
LB& 68780& 67157& 2.4\%& 67068& 0.1\%& 67031&    0.1\%& 2.5\%\\
TA& 46155& 44635& 3.3\%& 44541& 0.2\%& 44528& < 0.05\%& 3.5\%\\
RO& 45886& 44520& 3.0\%& 44467& 0.1\%& 44437&    0.1\%& 3.2\%\\
ZU& 45927& 44633& 2.8\%& 44584& 0.1\%& 44555&    0.1\%& 3.0\%\\
\hline
\textbf{total} & \textbf{414822} & \textbf{397999} & \textbf{4.1\%} & \textbf{397403} & \textbf{0.15\%} & \textbf{396809} & \textbf{0.15\%} & \textbf{4.4\%}\\
\hline
\end{tabular*}
\end{small}
\label{tab:data}
\end{table*}

\section{Algorithms}\label{sec:algorithms}

\subsection{ARMAX models}

Auto-Regressive (AR) and Moving-Average (MA) models with Exogenous Variables (ARMAX) describe a process as a linear combination of past measurements ($X_{t-i}$), past errors ($\epsilon_{t-i}$) and exogenous variables ($\boldsymbol{E_{t}}$, a vector which may include past values). If $p$ and $q$ are the orders of the AR and the MA terms respectively, the model is described by,

\begin{equation}
X_t = \alpha_{0} + \sum_{i=1}^{p} \alpha_{i} X_{t-i} + \sum_{j=1}^{q} \beta_{j} \epsilon_{t-i} + \boldsymbol{\gamma^{T}} \boldsymbol{E_{t}} + \epsilon_t,
\label{eq:armax}
\end{equation}

\noindent
where $\alpha_{0}$ is an independent term and $\epsilon_{t}$ is a deviation from the model at time $t$, assumed to be white Gaussian noise. The offset term ($\alpha_0$) allows to improve the modeling of processes of non zero mean. In this case $X_{t}$ (the forecast variable) will be the $k_{c}$ time series, which has a non-zero mean, therefore the offset term is useful to enhance the algorithm's performance. The exogenous variables $\boldsymbol{E_{t}}$ are also considered at time $t$. Using an ARMAX model to make forecasts implies finding the set of parameters $\alpha_{i}$, $\beta_{j}$ and $\gamma_{k}$ (one parameter for each exogenous variable), and then computing a step forward for some input at time $t$. ARMAX models are well-known as a prediction tool and are a natural generalization of ARMA models, which were popularized by \cite{Box-Jenkins-1970}.

\subsection{RLS filter}

Recursive Least Squares (RLS) is an optimization algorithm that recursively solves the minimization of a cost function depending on the weights $\boldsymbol{w_n}$,

\begin{equation}
C(\boldsymbol{w_n}) = \sum_{i=0}^{n} \lambda^{n-i} e^2(i),
\label{eq:Cwn}
\end{equation}

\noindent
where $e(i)$ is the forecasting error of observation $i$. If the lead time is $h$, then $e(i)= \boldsymbol{w_n^T z_{i-h}} - X_i $, being $\boldsymbol{z_{i-h}}$ a vector including all input variables and $X_i$ the target value. The factor $\lambda$ is called forgetting factor and when it is near 1, it resembles Least Squares Minimization while allowing the weights to adapt to the statistical changes of the $k_{c}$ time series. The algorithm presents some similarity to computing Least Squares in a moving window, being the main differences the exponentially decaying weights in the cost function and the fact that the computation is recursive. The mathematical generalization and resulting algorithm of the ARMAX-RLS framework for an arbitrary lead time $h$ is detailed in \cite{marchesoniacland}.

In traditional signal processing literature, the RLS algorithm is classified as an adaptive filter. Being adaptive makes historical data unnecessary, i.e. it avoids using train-test splits and fixed-weights, making the approach useful for operational context. Furthermore, statistical properties vary between seasons and even days (i.e. cloudy and clear-sky days), causing short-term adaptability to be a desirable property. This adaptability depends on the value of the forgetting factor $\lambda \in [0, 1]$. The coefficients of the filter are, as expected, more stable for a bigger value of $\lambda$. For large $\lambda$, the coefficients do not change much when a high variability period is found. On the other hand, when there is a clear sky, forecasts with large $\lambda$ consistently underestimate the target value due to the very low convergence rate. Using a smaller value of $\lambda$ makes convergence to the steady state of $k_{c}$ faster, yielding also unbiased estimates under clear sky conditions. However, the values of $\lambda$ closer to $1$ are the ones that achieve better numerical results in the long run. The loss function is a weighted sum of squared errors, making it convenient for any algorithm to `play safe' to some extent: an underestimating forecast under clear sky conditions will achieve less quadratic error when clouds appear. By inspecting this behavior the $\lambda$ value was set to $\lambda = 0.999$, as was previously used by \cite{david2016probabilistic}.

\section{Evaluation framework}\label{sec:performance}

\subsection{Metrics}\label{susec:metrics}

Results are presented in terms of the traditional Mean Bias Deviation (MBD) and Root Mean Squared Deviation (RMSD) metrics, as well as the Forecasting Skill (FS) metric. The MBD and RMSD definitions are,

\begin{equation}
\text{MBD}_h = \frac{1}{N} \sum_{i} \left[ \hat{y}_{h}(i) - y^{\text{ref}}(i+h) \right],
\end{equation}

\begin{equation}
\text{RMSD}_h = \sqrt{\frac{1}{N} \sum_{i} \left[ \hat{y}_{h}(i) - y^{\text{ref}}(i+h) \right]^{2}},
\end{equation}

\noindent
where $\hat{y}_h$ is the GHI forecast, $y^{\text{ref}}$ is the GHI measurement and $h$ is the forecast horizon. Their relative values, rMBD and rRMSD, are expressed as a percentage of the average irradiance value (see \Cref{tab:sites}). The MBD definition is such that a positive value means a forecasting overestimation and a negative value means a forecasting underestimation. The forecasting skill represents the gain of the forecasting RMSD with respect to the persistence procedure, and it is defined as,

\begin{equation}
\text{FS} = 1 - \frac{\text{RMSD}_{\text{m}}}{\text{RMSD}_{\text{p}}},
\end{equation}\label{eq:fs}

\noindent
where the subscripts `m' and `p' refer to the model and persistence respectively. The traditional persistence is calculated by setting $\hat{k}_{c} (t+h) = k_{c} (t)$, for every $h\geq1$, where $\hat{k}_{c} (t+h)$ is the predicted $k_{c}$ value. Then, the corresponding GHI is predicted by using the clear sky model estimates. Being a simple procedure, the persistence is then used as a benchmark to measure how good a forecast is: any additional complexity of the forecasting procedure should imply an improvement in comparison with the persistence to be worthwhile. One advantage of the FS metric over just reporting RMSD is that it makes algorithms that are tested on different geographical locations comparable, as the difficulty of forecasting at each location is implicit in the persistence's RMSD \citep{Coimbra-2013}. However, the choice of the benchmark over which to calculate the FS is arguable. An alternative to persistence is to persist in time the past observations' average, known as Smart Persistence (SP). For this work, the ``best'' SP is used as performance reference. This defines the forecasting skill metric that will be used later, defined exactly as in Eq~(\ref{eq:fs}), but using $\text{RMSD}_\text{sp}$ instead of $\text{RMSD}_\text{p}$. The procedure to obtain this best SP is detailed in the next subsection.

\subsection{Smart persistence}\label{susec:smart_persistence}

A slightly more complex variant of the classical persistence method is called ``smart persistence''. Persistence and smart persistence methods are defined in different ways in the literature. For example, \cite{VOYANT2018343} define ``persistence'' as persistence of the $\text{GHI}$ value and ``smart persistence'' as persistence of the $k_{c}$ index. \cite{david2016probabilistic} define ``smart persistence'' as the average of the last $h$ values of $k_{c}$, being $h$ the forecasting time horizon. In this work the definition is as follows: the general smart persistence implies persisting not the last observation nor the last $h$ observations, but the average of the last $n$, being $n$ an arbitrary amount of past samples that will be optimized for each time horizon. This is, $\hat{k}_{c} (t+h) = \frac{1}{n} \sum_{i=0}^{i=n-1} k_{c} (t-i)$. When $n=1$, the persistence is recovered, while when $n$ is large, the prediction approximates the climatology value ($n\gg1$). It is known that the shorter time horizons are better predicted by the classical persistence, while the longer time horizons are better predicted by the climatology value. In \cite{yang2019making} and \cite{yang2019standard} a linear combination of persistence and climatology is proposed as benchmark, after mathematically proving that the weights are directly related to the autocorrelation of the time series at the time horizon studied. Here it is argued that there exists an optimal $n_{h}^{*}$ value for each lead time $h$ (the value of $n$ that minimizes the prediction RSMD for that time horizon), which is not usually assessed nor presented.

\begin{figure*}[ht!]
\centering

\subfloat[rRMSD as a function of $h$.]
{\includegraphics[width=0.75\textwidth]{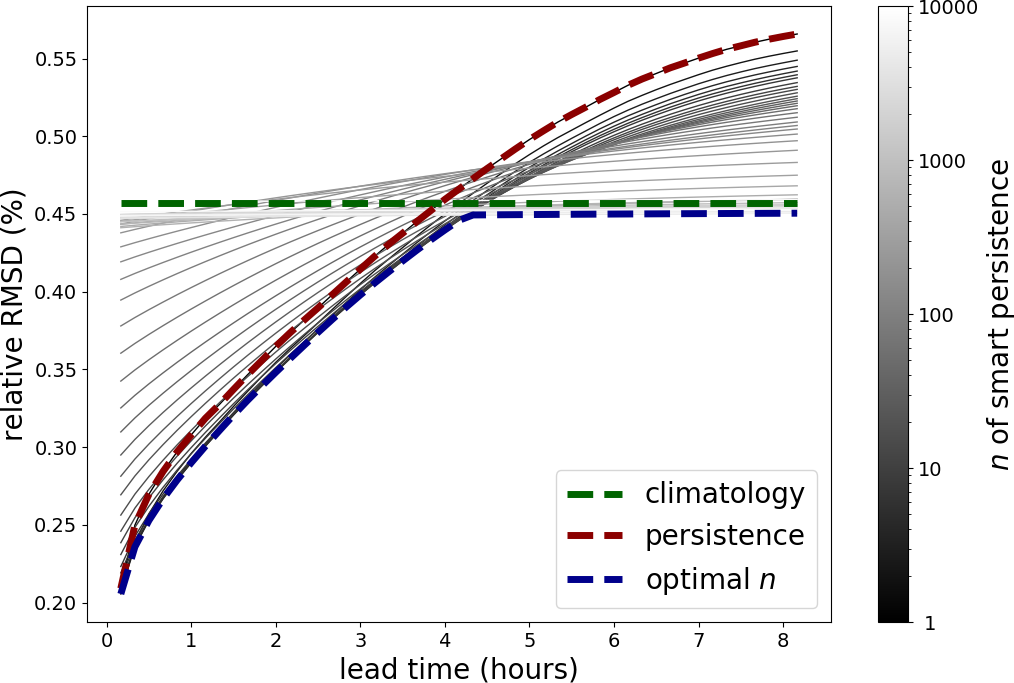}\label{fig:SP_rmsd_vs_lt}}

\subfloat[rRMSD as a function of $n$.]
{\includegraphics[width=0.75\textwidth]{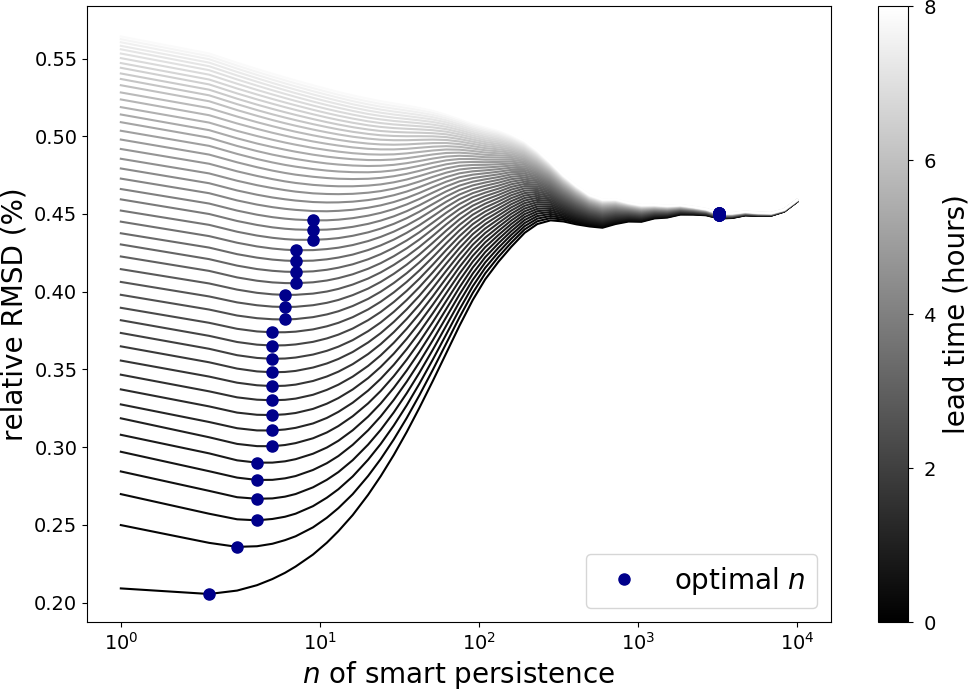}\label{fig:SP_rmsd_vs_n}}

\caption{Relative RMSD versus lead time for different values of smart persistence $n$.}
\label{fig:SP_rmsd}
\end{figure*}

To obtain the best SP, the $n_{h}^{*}$ were obtained for each site and time horizon. Performance evaluation of smart persistence is done via a grid search methodology: forecasts are made for every location, for every lead-time value up to $8$ hours ahead with 10 minutes granularity, and over different values of $n$. The results, depicted in \Cref{fig:SP_rmsd_vs_lt,fig:SP_rmsd_vs_n} for the average of all sites, show two things. First, the value of optimal $n$ increases with lead-time, but in a different manner as $n_{h}^{*}=h$. Second, there is a time horizon from which smart persistence forecasts with bounded $n$ are worse than climatology forecasts. This breakpoint in which big $n$ persistence (\textit{pseudo} climatology) is better than small $n$ persistence, is best observed in \Cref{fig:SP_rmsd_vs_n}. It can be seen that smart persistence optimum $n$ increases slowly (note that the $x$-axis is logarithmic) with forecasting lead time until the breakpoint is reached. In the breakpoint, the forecasting lead time is large enough to make the information contained in the recent past samples less valuable than the historical aggregate. The breakpoint for the region under study is located at $4$ hours and $20$ minutes (sites' average). For time horizons longer than this point it makes little sense to benchmark with any kind of smart persistence, as the simple climatology value will be better.

The procedure for obtaining the ``best smart persistence'' can be viewed as computing the RMSD curve (vs $h$) for each possible value of $n$, and then taking the bottom envelope of the curves as the $\text{RMSD}_{\text{sp}}$. \Cref{fig:SP_rmsd_vs_lt} shows in gray scale the site-averaged RMSD curves with varying $n$, and three special curves are identified: the classical persistence in red, the climatology in green and the best persistence performance in blue. This RMSD bottom envelope, for each station, is the benchmark used in this work to define the FS from Eq~(\ref{eq:fs}).

In a real case scenario, the $n_{h}^{*}$ values (over past and future) are unknown. Optima obtained from historical data are not guaranteed to keep being optimal in the future, although the likelihood of this happening grows with historical data size. The same thing happens when using climatology as a benchmark. It is noted here that this subtle case of data snooping is harmless because persistence models are only used as benchmarks instead of operationally. The procedures presented here and in \cite{yang2019standard} are of different nature, being the present benchmark more computationally expensive. However, this optimal smart persistence benchmark is computed only one time as a standalone calculation for the purpose of performance assessment, hence it does not represent a limitation for operational systems. Future work should include the comparison of both methodologies, addressing different climates and solar variability sites. 

\section{Results}\label{sec:results}

This section presents the performance results of the forecasting algorithms. The focus is on understanding and quantifying the performance gain of adding averaged satellite albedo to the ARMA-RLS baseline model that only uses ground measurements. \Cref{susec:RLS_measurements} introduces the results for the baseline model, providing a discussion on model selection and showing that model fine-tuning is of little utility. \Cref{susec:RLS_cloudiness} presents the core results of this article, addressing the inclusion of satellite albedo into the auto-regressive framework. The results in this section are presented via graphical aids. Nevertheless, the complete quantitative results are provided in Tables in \Cref{apendice}.

\newpage

\subsection{Endogenous RLS filter}\label{susec:RLS_measurements}

\begin{figure*}[ht!]
\centering

\subfloat[Forecast horizon: 10 minutes ($h=1$).]
{\includegraphics[width=0.46\textwidth]{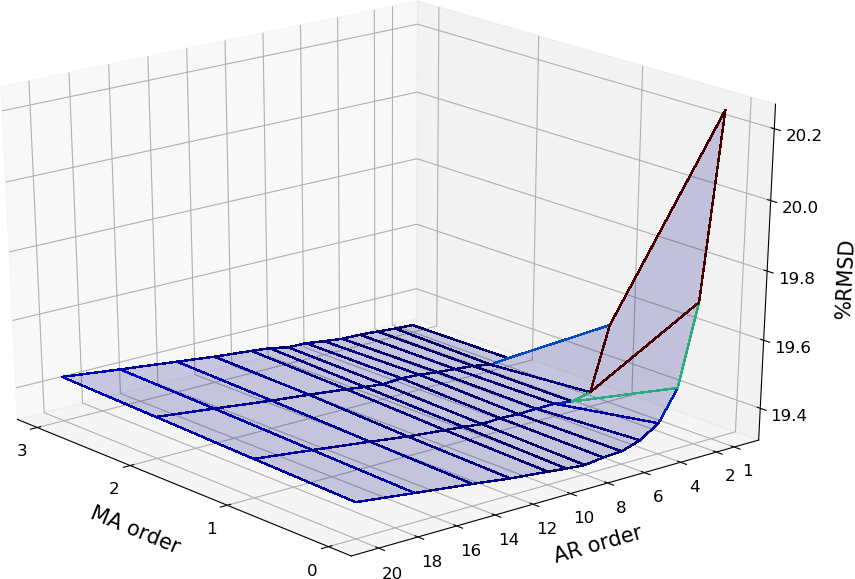}\label{fig:surf_lt_1}}
\hspace{2mm}
\subfloat[Forecast horizon: 90 minutes ($h=9$).]
{\includegraphics[width=0.46\textwidth]{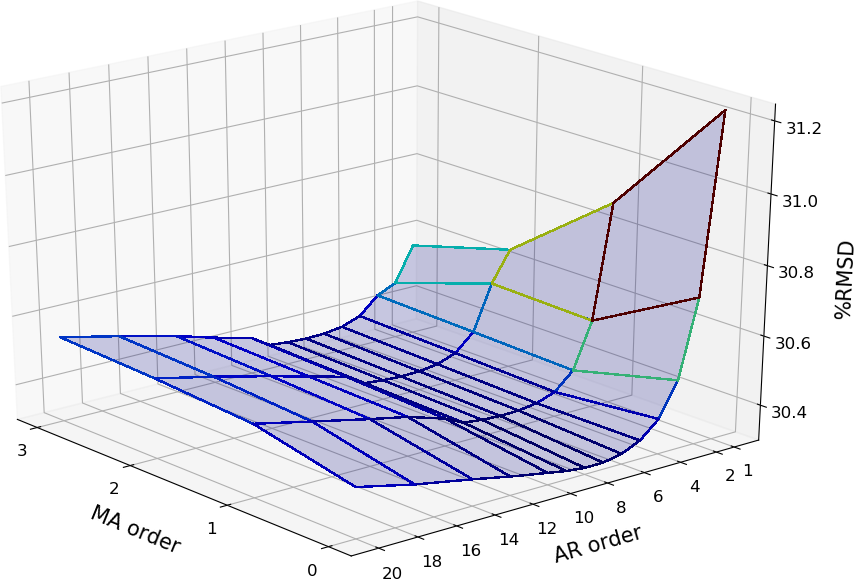}\label{fig:surf_lt_9}}

\subfloat[Forecast horizon: 240 minutes ($h=24$).]
{\includegraphics[width=0.46\textwidth]{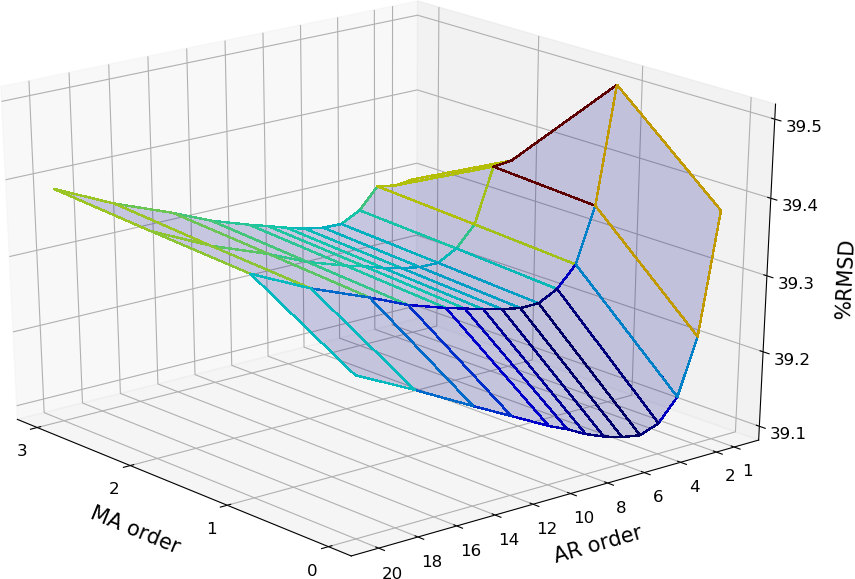}\label{fig:surf_lt_24}}
\hspace{2mm}
\subfloat[Average rRMSD anomalies over all lead times.]
{\includegraphics[width=0.46\textwidth]{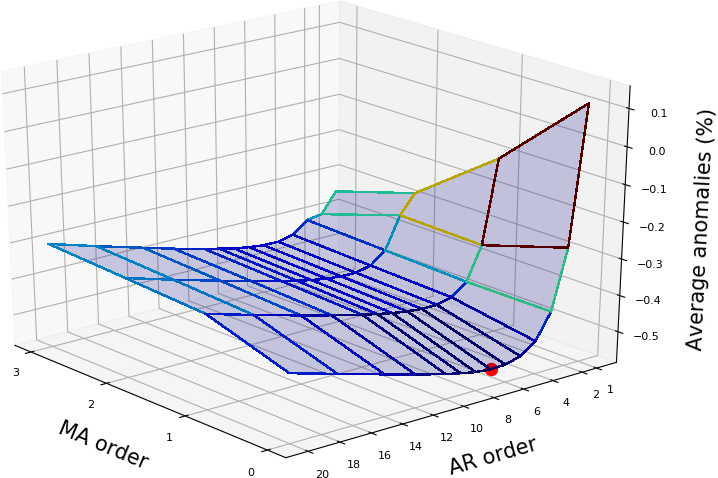}\label{fig:surf_avg}}

\caption{Relative RMSD analysis for the different $(p, q)$ parameters of the endogenous ARMA-RLS filter.}
\label{fig:rRMSD_analysis_pq}
\end{figure*}

\Cref{fig:surf_lt_1,fig:surf_lt_9,fig:surf_lt_24} show the rRMSD for each $p$ (number of AR terms) and $q$ (number of MA terms) averaged over all locations. It is observable in these examples that with $q=0$ and $3 \leq p \leq 8$ the performance is near the optimal one. For the shorter time horizons ($h=1$, 10 minutes, \Cref{fig:surf_lt_1}), after a certain value of $p$ and $q$, i.e. for $p\geq3$ and $\forall q$, the performance is rather similar, with rRMSD variations below $0.1$\%. The rRMSD span over all $p$ and $q$ values is $\simeq0.8$\%, which is not very high. For the longer time horizons ($h=24$, 4 hours,  \Cref{fig:surf_lt_24}), the surface seems to favor $q = 0$ more clearly, but the performance difference over different values of $q$ and $p$ near the optimum is below $0.2$\% (negligible, see the plot's $z$-axis scale). It is clear that for large time horizons, there is less performance gain by tuning the $p$ and $q$ values for each lead time. Intermediate lead times stand between both situations. Hence, for all time horizons, the gain obtained by fine-tuning $p$ and $q$ is below $0.8$\% of rRMSD. Furthermore, we will shortly show that if a fixed pair of these values is intelligently set, there is no global significant performance gain in optimizing $(p, q)$ for each lead time.

To find a good global model with fixed $(p, q)$ parameters, the procedure is as follows: we calculated all the rRMSD surfaces averaged over the locations, subtracted the mean of the rRMSD at each lead time, and averaged the result. This procedure obtains the mean rRMSD anomalies surface, whose minimum is the best fixed operation point. This procedure avoids giving more importance to one specific lead time. The result is shown in \Cref{fig:surf_avg} and the minimum is located at $(6, 0)$. This is very close to the ad-hoc $(5,0)$ model that we analyzed in a previous work \citep{marchesoniacland}. In that preliminary work, we observed that the performance of an arbitrary $(p=5, q=0)$ model was indistinguishable from the best error achievable with any bounded combination of $p$ and $q$ (the bounds were $p\leq 10$ and $q\leq4$). For simplicity, and as there is a negligible gain in using one or the other, we will use in the following $p=5$ for the analysis.

\begin{figure*}[ht!]
\centering

\subfloat[Using regular persistence.]
{\includegraphics[width=0.46\textwidth]{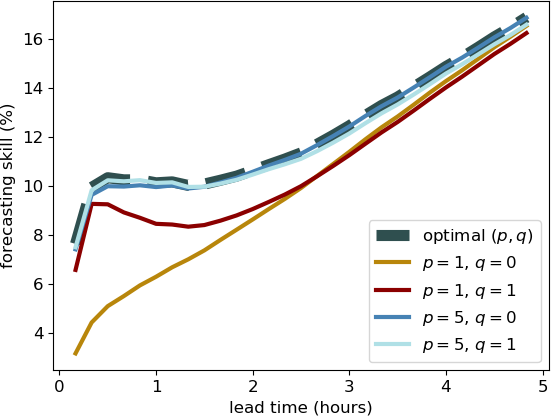}\label{fig:fs_rls}}
\hspace{3mm}
\subfloat[Using the benchmark optimal smart persistence.]
{\includegraphics[width=0.46\textwidth]{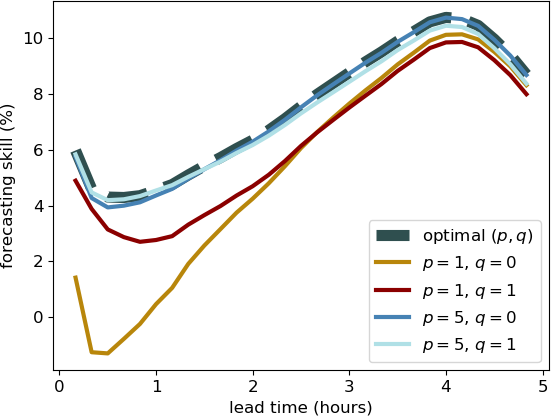}\label{fig:fs_star_rls}}

\caption{Forecasting Skill of the ARMA-RLS filter with different parameters in comparison with optimal choice.}
\label{fig:optimal_choice}
\end{figure*}

\Cref{fig:fs_rls,fig:fs_star_rls} show the forecasting skill of different $(p, q)$ models in comparison with the optimal model, i.e. the model that uses the optimal $p$ and $q$ values for each time horizon. For easy comparison with other works and to visualize the effect of using different persistence procedures, \Cref{fig:fs_rls,fig:fs_star_rls} show the FS using the regular persistence and the optimal smart persistence, respectively. It is to be noted the different span ($y$ axis) and the different behavior for shorter lead times (up to 1 hour ahead) and longer lead times (for $4$-$5$ hours ahead, when the pseudo climatology is the best benchmark), where the concavities are different. The performance when using $(5, 0)$ and $(5, 1)$ remains very close to each other and to the optimal $(p, q)$ model. This in fact shows that using a fixed well-selected pair of $(p, q)$ obtains essentially the same performance as the optimal choice and that fine-tuning the ARMA-RLS filter is futile. For short time horizons, the effect of adding a MA term ($q$) is positive for $p=1$, but is insignificant for $p=5$. For long time horizons, adding a MA term tends to slightly downgrade the performance.

\subsection{RLS filter including satellite albedo}\label{susec:RLS_cloudiness}

Including satellite albedo data improves the models' performance compared to using only ground measurements, as shown in this subsection. The models used here will only include AR terms, as the difference is insignificant for $p \geq 3$. Henceforth, ``lags'' will be used to refer to past satellite observations, i.e. albedo observations previous to time $(t)$. Also, the performances will be expressed only in terms of the FS metric using as benchmark the optimal smart persistence in order to avoid redundancy. The cases with $p=5$ and $p=1$ and no satellite input are included in the following figures as a performance reference, and the FS curves are the same of \Cref{fig:fs_star_rls} for the $(5,0)$ and $(1,0)$ models, respectively.

\begin{figure*}[ht!]
\centering

\subfloat[With varying $p$.]
{\includegraphics[width=0.48\textwidth]{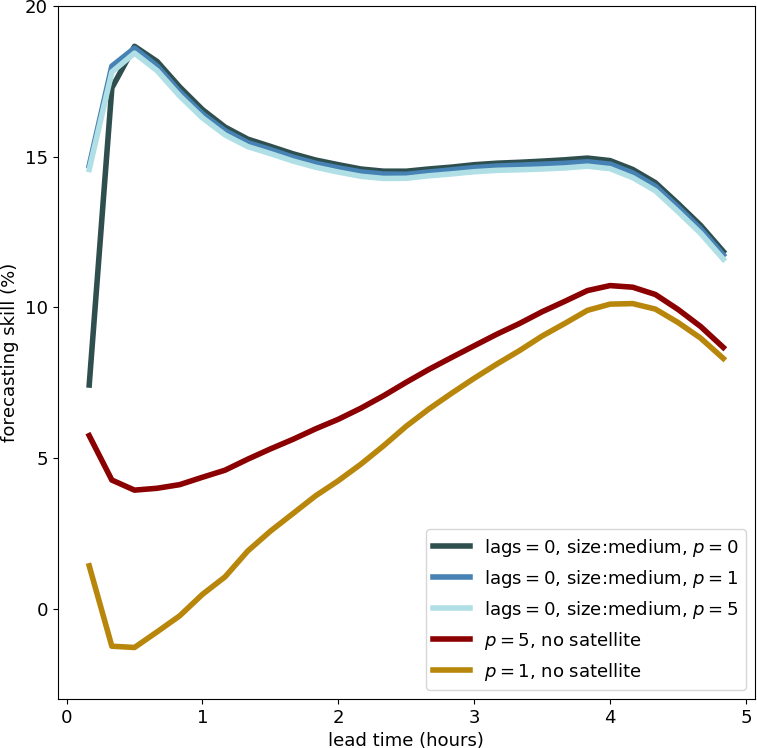}\label{fig:sat_fs_star_vs_p}}
\hspace{3mm}
\subfloat[With varying satellite lags.]
{\includegraphics[width=0.48\textwidth]{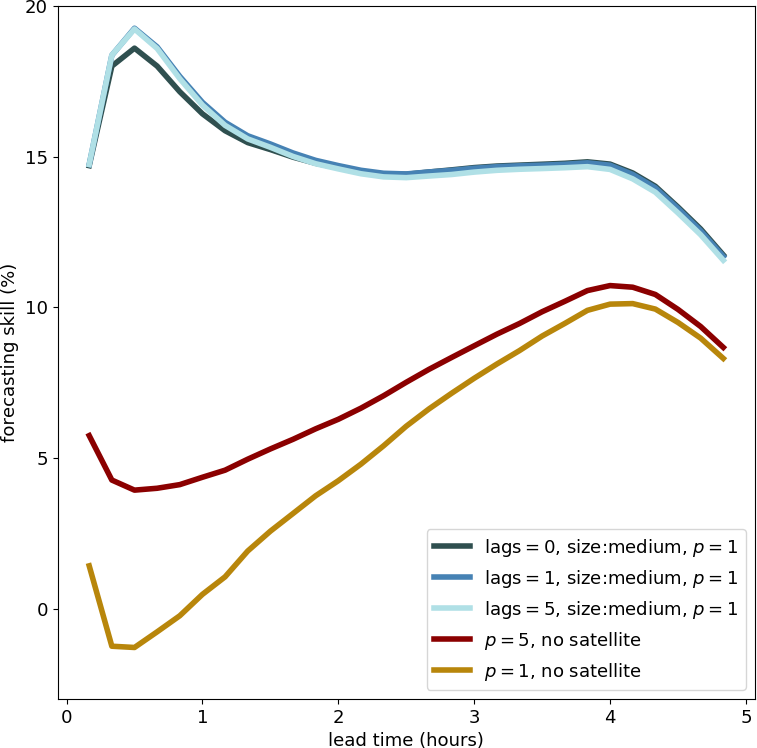}\label{fig:sat_fs_star_vs_lags}}

\caption{Forecasting Skill for ARMA-RLS filter using satellite albedo.}
\label{fig:sat_fs_star_1}
\end{figure*}

The effect of the number of AR terms $p$ on the performance when including a single value of satellite albedo data is presented in \Cref{fig:sat_fs_star_vs_p}. The pixel size used here is medium (see \Cref{susec:satellite}). The addition of satellite albedo enhances performance significantly for all lead times. Peak performance is obtained at 30 minutes ahead ($\text{FS} \simeq +18.5$\%), being a $\times~4$ improvement over the ARMA $(5,0)$ model for that time horizon. In general, it can be seen that the higher impact of adding satellite albedo is at the shorter time horizons, i.e. during the first hour ahead, but then a remnant improvement persists for longer lead times, declining during the last forecast hour ($4$-$5$ hours ahead) in the same manner as the rest of the models. Furthermore, \Cref{fig:sat_fs_star_vs_p} shows that increasing $p$ ceases to be useful when satellite data is present (see $p=1$ and $p=5$ cases). Taking this to the edge and using no ground data in the algorithm's input ($p=0$) only implies sacrificing significant performance on the 10 minutes lead time, as the FS is almost the same for larger lead times. This does not mean that ground measurements are unnecessary: they are used to generate the error signal that is fed back into the RLS algorithm. Therefore, this does not mean that the $p=0$ algorithm is only running on satellite information. The use of solar satellite estimates to completely replace the measurement signal is left as future work, but recent studies suggest this may be possible without a significant performance reduction \citep{yang2019can}.

Another experiment was made: adding lags on the satellite cloudiness data. It should be noted that, as the 10-minutes satellite data series is obtained via interpolation from a smaller time resolution, there is some degree of redundancy in this information. Adding lags can be seen as a type of time-averaging, as a set of weights $\{ \gamma_{k} \}$ will be assigned to each past data-point. The value $p=1$ was used in this test as there is no significant improvement by using $p=5$ when satellite albedo is also used (as seen in \Cref{fig:sat_fs_star_vs_p}). The analysis is shown in \Cref{fig:sat_fs_star_vs_lags}'s blued curves. A little performance improvement on the peak of the FS curve is found, of around $1$\% of FS. For larger time horizons the effect is negligible. This happens for both $\text{lags}=1$ and $\text{lags}=5$, showing a behavior similar to that of the $p$ value: there is no extra value in adding more satellite lags than $\text{lags}=1$. In \Cref{fig:sat_fs_star_vs_lags} the curve with one satellite lag is indistinguishable from the one with five satellite lags.

\newpage

\begin{figure}[ht!]
\centering

\subfloat[With varying pixel size.]
{\includegraphics[width=0.48\textwidth]{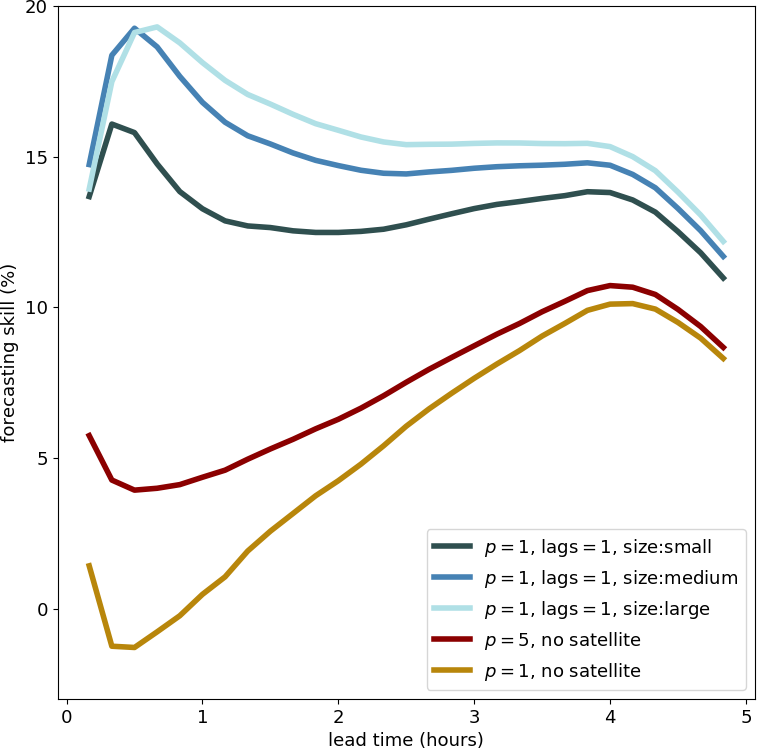}\label{fig:sat_fs_star_vs_size}}
\hspace{3mm}
\subfloat[With varying satellite lags and pixel size.]
{\includegraphics[width=0.48\textwidth]{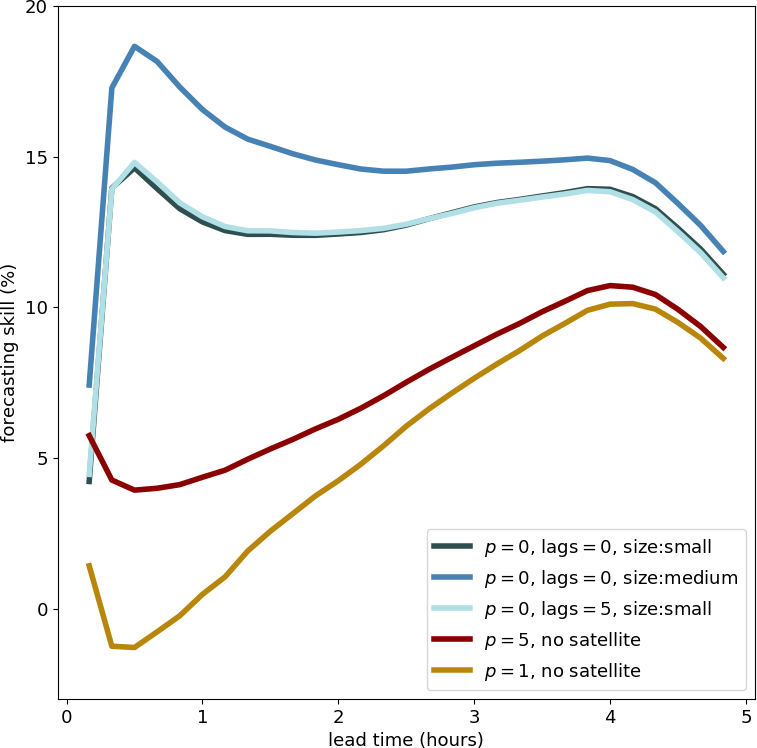}\label{fig:ergodicity}}

\caption{Forecasting Skill for ARMA-RLS filter using satellite albedo.}
\label{fig:sat_fs_star_2}
\end{figure}

The third analysis is about the impact of the spatial window in which the values of satellite albedo are averaged. $p=1$ and one lag on satellite data are used in this case, in order to quantify the impact of the window size in the best model inspected so far. \Cref{fig:sat_fs_star_vs_size} shows the models' performance when using the small, medium and big cell size (defined in \Cref{susec:satellite}). It is observed that larger cell sizes are preferred. A significant performance improvement is observed when using a medium cell size in comparison with a small cell size, especially up to $\simeq2$ hours ahead. The bigger cell inspected here is the one which provides better performance, being similar to that of the medium cell size up to the 30 minutes time horizon, but showing an improvement for longer ones. Another interesting observation is the location of the peak: using a larger spatial window implies moving the peak in the direction of larger lead times. Note also that the concavity of all curves that include satellite data is negative in the shorter lead times, denoting a relative advantage over forecasts methods that do not include satellite data in these time horizons: spatially averaged satellite albedo effectively improves the forecast in the first forecast hour and has a positive effect over all time horizons.

The last test is shown in \Cref{fig:ergodicity} and refers to the absence or presence of the ``ergodicity'' property of satellite albedo images for solar forecasting. This means analyzing whether time-averaging and spatial averaging are interchangeable or not \emph{for forecasting purposes}. In absence of any other information, it was tested if using satellite lags (i.e. weighted time averaging) is the same as using a spatially averaged satellite input, without lags. The former is tested by using $p=0$, $\text{lags}=5$ and a small window size, and the latter is tested by using $p=0$, $\text{lags}=0$ and a medium window size. The baseline level of $p=0$, $\text{lags}=0$ and a small window size is also given, as a reference. It can be seen that the model including satellite lags shows almost no improvement from the baseline level. However, the model using a medium satellite window size reaches a significant improvement. Hence, it is clear that including spatially averaged satellite information is more useful than using time-averaged satellite information. In other words, weighted time averaging is not equivalent to spatial averaging over satellite data in terms of forecasting performance.

Summing up, the simpler best performing ARMAX-RLS model found here is that with $p=1$, one lag in satellite data and a large spatial window ($37~\text{km}\times31~\text{km}$), followed closely by the medium spatial window ($19~\text{km}\times16~\text{km}$). It has a peak FS slightly above $+19$\% at 40 minutes ahead ($h=4$) and a better performance than the previously tested models for $h\geq4$. From 10 to 30 minutes ahead its performance is similar to that of the same model but using a medium satellite albedo window, being $\simeq+1$\% lower in the first two lead times. Its FS is $\simeq+16\%$ for most lead times between $2$ and $4$ hours, showing the typical downgrade in the last hour, as seen in previously tested models. The FS of all the tested models are positive, even using the optimal smart persistence as benchmark for its calculation, with the only exception of the model using $p=1$ but no satellite data. Based on this, it can be argued that simple ARMA-RLS models based only on ground measurements, like the $(5,0)$ model, could be used as a more exigent performance benchmark for solar forecasting methods. Further studies on this topic should include different climates.

The closest work in the literature is the one of \cite{Dambreville-2014}. The comparison can be made in terms of the regular FS (provided in this work in \Cref{apendice}), but it is not straightforward due to different time scales and locations under study. \citeauthor{Dambreville-2014} use a 15 minutes times basis (the MSG satellite time resolution) and tested their ideas using ground measurements from an urban BSRN site in Paris, France (PAL station, SIRTA Observatory). The short-term variability ($\sigma$) of the site is not provided, but one can use the absolute RMSD of the regular persistence as an indication of the sites' similarity. \Cref{tab:comparison} provides the comparison between both works. We use the 10 minutes time basis in this work, so the 15 and 45 minutes values were linearly interpolated from \Cref{tab:MBD_RMS_ground,tab:FS_ground,tab:FS_satellite} in order to make the comparison possible. The AR(5) model stands for the ARMA model with $p=5$ and $q=0$. The AST method uses as input the $3\times3$ fixed pixels centered at the site's location and the AST2 method uses intercorrelation maps to decide which pixels are more useful as input for each time horizon. For more details on these last two models, please see \cite{Dambreville-2014}. The convention in \Cref{tab:comparison} for the satellite models of this work is SAT($p$, lags) and the used pixel size is medium. It is observed that in the Paris site the regular persistence's RMSD starts lower than in our region, but increases more quickly, hence the solar variability regimen is not the same, although rather similar. The satellite models' FS are of similar order, but there is an important difference in their behavior for both works: for \citeauthor{Dambreville-2014} they increase with the time horizon while for the present work they have a maximum around 30 minutes. As the same behavior is observed with the AR(5) model, which does not use satellite information, we think this phenomenon is explained by the different behavior of the regular persistence. In fact, the AR(5) model is included here as a reference between both works that does not take into account the way that satellite information is used (which is different). To isolate the contribution of the satellite input it is possible to take the FS difference of each model with respect to the AR(5) model, also shown in \Cref{tab:comparison} as `gain'. It is observed that the AST2, SAT(1,0) and SAT(1,1) models have similar gains, around $+9$-$14$\%, while for the AST model the gain is lower. The AST2 model presents a slightly better gain of $\simeq1$\% than that of the SAT models for the first time horizon considered (15 minutes), but this gain then decreases monotonically. On the other hand, the SAT models have a maximum gain at $30$ minutes ahead of $+13$-$14$\%, outperforming in $\simeq+3$\% the AST2 model between $30$ and $60$ minutes ahead. The gain with respect to the AR(5) model also allows to visualize better the slight improvement obtained when including satellite lags (SAT(1,1) model vs SAT(1,0) model), of around $+0.5$\% for these time horizons. We conclude that the proposals of the present work are a simple and effective alternative to include satellite information into solar forecasting methods and its performance is competitive with other more complex approaches.

\begin{table*}[ht!]
\renewcommand{\arraystretch}{1.1}
\centering
\caption{Comparison between the work of Dambreville et al. (2014) and the present work. The information from Dambreville et al. was taken from the Table 1 of their work. The 15 and 45 minutes values for this work were obtained via linear interpolation of the 10-minutes metrics. The 30 and 60 minutes were taken directly from the 10 minutes evaluation.}\label{tab:comparison}
\begin{small}
\begin{tabular*}{\textwidth}{c@{\extracolsep{\fill}}C{2.5cm}C{1.5cm}C{1.5cm}C{1.5cm}C{1.5cm}C{1.5cm}}
\hline
\multicolumn{7}{c}{\textbf{Dambreville et al. (2014)}}\\
\hline
\textbf{lead}&\textbf{RMSD}&\multicolumn{3}{c}{\textbf{FS (\%)}}& \multicolumn{2}{c}{\textbf{gain vs AR(5)}}\\
\textbf{time}&\textbf{persist.}& \textbf{AR(5)}& \textbf{AST}& \textbf{AST2}& \textbf{AST}& \textbf{AST2}\\
\cline{1-1}\cline{2-2}\cline{3-5}\cline{6-7}
15~mins.&  $94$~W/m$^2$&  $+8.5$& $+17.0$& $+20.2$& $+8.5$& $+11.7$\\
30~mins.& $118$~W/m$^2$& $+15.3$& $+20.3$& $+26.3$& $+5.1$& $+11.0$\\
45~mins.& $130$~W/m$^2$& $+17.7$& $+20.8$& $+27.7$& $+3.1$& $+10.0$\\
60~mins.& $140$~W/m$^2$& $+20.7$& $+22.9$& $+29.3$& $+2.1$&  $+8.6$\\
\hline
\hline
\multicolumn{7}{c}{\textbf{This work}}\\
\hline
\textbf{lead}&\textbf{RMSD}&\multicolumn{3}{c}{\textbf{FS (\%)}}&
\multicolumn{2}{c}{\textbf{gain vs AR(5)}}\\
\textbf{time}&\textbf{persist.}& \textbf{AR(5)}& \textbf{SAT(1,0)}& \textbf{SAT(1,1)}& \textbf{SAT(1,0)}& \textbf{SAT(1,1)}\\
\cline{1-1}\cline{2-2}\cline{3-5}\cline{6-7}
15~mins.& $102$~W/m$^2$&  $+8.5$& $+19.4$& $+19.6$& $+10.9$& $+11.1$\\
30~mins.& $120$~W/m$^2$& $+10.2$& $+23.7$& $+24.4$& $+13.5$& $+14.2$\\
45~mins.& $129$~W/m$^2$& $+10.0$& $+22.7$& $+23.2$& $+12.7$& $+13.2$\\
60~mins.& $137$~W/m$^2$&  $+9.9$& $+21.3$& $+21.7$& $+11.4$& $+11.8$\\
\hline
\end{tabular*}
\end{small}
\end{table*}

\newpage

\section{Conclusions}\label{sec:conclusions}

Three things were done in this work: an analysis of smart persistence obtaining a novel benchmarking reference, a revisit on the optimal order of an ARMA model embedded in a RLS algorithm, and more importantly, a study of the impact of satellite data and its time and spatial averaging on the performance of solar forecasts made through a RLS filter approach. The resulting model is a simple alternative for including satellite information into solar forecasts and outperforms the best smart persistence, having a similar performance than other more sophisticated ways of using satellite data.

On the smart persistence analysis it was shown that the optimal value of $n$ depends on the lead time considered. As expected, this optimal value of $n$ grows with the forecasting horizon, but it is never equal to $1$, i.e. the regular persistence. Furthermore, there is a breakpoint at a lead time of approximately $4$ hours, in which comparing with (smart) persistence is not useful anymore, and comparison with climatology should be made. The optimal value of $n$ for each time horizon defines a best smart persistence, which is used as performance benchmark.

The RLS filter is a flexible algorithm that does not need train-validation-test splits and is suitable for formulating ARMAX models, so it was used here to assess the performance impact of including satellite information to baseline models that only use ground measurements. When ignoring the exogenous part (the satellite input), optimal orders can be found via a grid search and, for solar irradiance data, an ARMA model with fixed $3 \leq p \leq 8$ and $q=0$ performs almost optimally. In fact, we found that there is no value in finding the optimal $p$ and $q$ values for each time horizon, as the fixed parameters filter provides performance results indistinguishable from the optimal ones.

There are five remarks to be made regarding the inclusion of satellite data. (I) including satellite data removes the importance of ground measurements \emph{as inputs}, restricting their usefulness to the first 10-minutes time horizon (although, in the present formulation, they are still needed to feedback the error signal to the RLS algorithm). (II) Adding lags in satellite data only achieves little improvements for $30$-$40$ minutes ahead, on the FS curve peak. (III) Enlarging the spatial averaging window enhances performance: performance improvement increases quickly with the window size, but after a certain size, the improvement is restricted to the larger time horizons (higher than 1 hour ahead) at a cost of losing little performance in the first two time horizons. (IV) Enlarging the spatial averaging window moves the maximum of the FS curve in the direction of larger lead times. (V) As time-averages (including lags) does not yield the same performance improvement than spatial-averages, ergodicity does not seems to be a property of satellite albedo as input for solar forecasting.

The results presented here are valid, a priori, only for intermediate solar variability sites and regions with similar climates to the target region. Further research is required to fully understand the presented ideas for solar forecasting in various context, namely, making a simple use of satellite images. Testing these ideas for other sites and climates in the world, at least, accounting for the GOES-East satellite coverage, is part of our current work.

\section*{Acknowledgments}

The authors thank the SONDA network of the INPE (Brazil) for providing the S\~{a}o Martinho da Serra data set. They also gratefully acknowledge the financial support given by Uruguay's National Research and Innovation Agency (ANII) under the FSE-ANII-2016-131799 grant.

\appendix
\renewcommand*{\thesection}{\Alph{section}}
\section{Detailed performance metrics}\label{apendice}

In this appendix we provide the detailed performance results, including the rMBD, rRMSD and FS metrics, the latter based on the regular persistence and the optimal smart persistence, for easy comparison with other works. \Cref{tab:MBD_RMS_ground,tab:FS_ground} present the results for the endogenous models showed in \Cref{fig:optimal_choice}. \Cref{tab:MBD_RMS_ground} also provides the rMBD and rRMSD metrics for both persistence methods. \Cref{tab:MBD_RMS_satellite,tab:FS_satellite} present the results for all the tested models that include space-averaged satellite albedo, that were analyzed in \Cref{fig:sat_fs_star_1,fig:sat_fs_star_2}.

\newpage

\begin{table*}[ht!]
\renewcommand{\arraystretch}{1.1}
\centering
\caption{Relative MBD and RMSD metrics for the persistence and the models that only use ground measurements.}\label{tab:MBD_RMS_ground}
\begin{small}
\begin{tabular*}{\textwidth}{c@{\extracolsep{\fill}}C{1.4cm}C{1.4cm}C{1.4cm}C{1.4cm}C{1.4cm}C{1.4cm}}
\cline{1-5}
\multicolumn{5}{c}{\textbf{ARMA-RLS model specification}}&&\\
\cline{1-5}
$\mathbf{p}$& 1& 1& 5& 5& \textbf{regular}& \textbf{smart}\\
$\mathbf{q}$& 0& 1& 0& 1& \textbf{persist.}& \textbf{persist.}\\
\hline
\textbf{lead time}&\multicolumn{6}{c}{\textbf{relative MBD (\%)}}\\
\cline{1-1}\cline{2-7}
 10~mins& $-0.3$& $-0.2$& $-0.2$& $-0.2$& $-0.1$& $-0.1$\\
 20~mins& $-0.4$& $-0.4$& $-0.4$& $-0.4$& $-0.1$& $-0.2$\\
 30~mins& $-0.5$& $-0.5$& $-0.5$& $-0.5$& $-0.2$& $-0.2$\\
 40~mins& $-0.6$& $-0.6$& $-0.6$& $-0.6$& $-0.2$& $-0.3$\\
 50~mins& $-0.7$& $-0.7$& $-0.7$& $-0.7$& $-0.3$& $-0.4$\\
 60~mins& $-0.8$& $-0.8$& $-0.8$& $-0.8$& $-0.3$& $-0.5$\\
 90~mins& $-1.0$& $-1.0$& $-1.0$& $-1.0$& $-0.5$& $-0.7$\\
120~mins& $-1.2$& $-1.2$& $-1.2$& $-1.2$& $-0.7$& $-0.9$\\
150~mins& $-1.3$& $-1.3$& $-1.3$& $-1.3$& $-0.9$& $-1.1$\\
180~mins& $-1.4$& $-1.4$& $-1.4$& $-1.4$& $-1.1$& $-1.2$\\
220~mins& $-1.5$& $-1.5$& $-1.6$& $-1.5$& $-1.3$& $-1.3$\\
260~mins& $-1.6$& $-1.5$& $-1.6$& $-1.6$& $-1.4$& $-1.9$\\
\hline
\textbf{lead time}&\multicolumn{6}{c}{\textbf{relative RMSD (\%)}}\\
\cline{1-1}\cline{2-7}
 10~mins& $20.3$& $19.5$& $19.4$& $19.3$& $20.9$& $20.5$\\
 20~mins& $23.9$& $22.7$& $22.6$& $22.5$& $25.0$& $23.6$\\
 30~mins& $25.6$& $24.5$& $24.3$& $24.2$& $27.0$& $25.3$\\
 40~mins& $26.9$& $25.9$& $25.6$& $25.5$& $28.4$& $26.7$\\
 50~mins& $27.9$& $27.1$& $26.7$& $26.7$& $29.7$& $27.9$\\
 60~mins& $28.9$& $28.2$& $27.7$& $27.7$& $30.8$& $29.0$\\
 90~mins& $31.2$& $30.9$& $30.4$& $30.4$& $33.7$& $32.1$\\
120~mins& $33.3$& $33.2$& $32.6$& $32.7$& $36.5$& $34.8$\\
150~mins& $35.1$& $35.1$& $34.6$& $34.6$& $39.0$& $37.4$\\
180~mins& $36.7$& $36.8$& $36.3$& $36.4$& $41.5$& $39.8$\\
220~mins& $38.6$& $38.7$& $38.3$& $38.4$& $44.5$& $42.6$\\
260~mins& $40.1$& $40.2$& $39.9$& $40.0$& $47.3$& $44.5$\\
\hline
\end{tabular*}
\end{small}
\end{table*}

\newpage

\begin{table*}[ht!]
\renewcommand{\arraystretch}{1.1}
\centering
\caption{Forecasting skill of the models that only use ground measurements.}\label{tab:FS_ground}
\begin{small}
\begin{tabular*}{0.7\textwidth}{c@{\extracolsep{\fill}}C{1.5cm}C{1.5cm}C{1.5cm}C{1.5cm}}
\hline
\multicolumn{5}{c}{\textbf{ARMA-RLS model specification}}\\
\hline
$\mathbf{p}$& 1& 1& 5& 5\\
$\mathbf{q}$& 0& 1& 0& 1\\
\hline
\textbf{lead time}& \multicolumn{4}{c}{\textbf{FS (\%) using the regular persistence}}\\
\cline{1-1}\cline{2-5}
 10~mins&  $+3.2$&  $+6.6$&  $+7.4$&  $+7.5$\\
 20~mins&  $+4.4$&  $+9.2$&  $+9.6$&  $+9.8$\\
 30~mins&  $+5.1$&  $+9.2$& $+10.0$& $+10.2$\\
 40~mins&  $+5.5$&  $+8.9$& $+10.0$& $+10.2$\\
 50~mins&  $+5.9$&  $+8.7$& $+10.0$& $+10.2$\\
 60~mins&  $+6.3$&  $+8.4$&  $+9.9$& $+10.1$\\
 90~mins&  $+7.4$&  $+8.4$& $+10.0$& $+10.0$\\
120~mins&  $+8.6$&  $+9.0$& $+10.6$& $+10.5$\\
150~mins&  $+9.9$& $+10.0$& $+11.3$& $+11.1$\\
180~mins& $+11.4$& $+11.2$& $+12.4$& $+12.1$\\
220~mins& $+13.3$& $+13.1$& $+14.0$& $+13.7$\\
260~mins& $+15.2$& $+14.9$& $+15.6$& $+15.4$\\
\hline
\textbf{lead time}& \multicolumn{4}{c}{\textbf{FS (\%) using the optimal smart persistence}}\\
\cline{1-1}\cline{2-5}
 10~mins&  $+1.4$& $+4.9$&  $+5.7$&  $+5.8$\\
 20~mins&  $-1.2$& $+3.9$&  $+4.3$&  $+4.5$\\
 30~mins&  $-1.3$& $+3.1$&  $+3.9$&  $+4.2$\\
 40~mins&  $-0.8$& $+2.9$&  $+4.0$&  $+4.2$\\
 50~mins&  $-0.3$& $+2.7$&  $+4.1$&  $+4.3$\\
 60~mins&  $+0.5$& $+2.8$&  $+4.3$&  $+4.6$\\
 90~mins&  $+2.6$& $+3.7$&  $+5.3$&  $+5.3$\\
120~mins&  $+4.3$& $+4.7$&  $+6.3$&  $+6.2$\\
150~mins&  $+6.0$& $+6.1$&  $+7.5$&  $+7.3$\\
180~mins&  $+7.6$& $+7.5$&  $+8.7$&  $+8.4$\\
220~mins&  $+9.5$& $+9.2$& $+10.2$&  $+9.9$\\
260~mins& $+10.0$& $+9.7$& $+10.4$& $+10.1$\\
\hline
\end{tabular*}
\end{small}
\end{table*}

\newpage

\begin{table*}[ht!]
\renewcommand{\arraystretch}{1.1}
\centering
\caption{Relative MBD and RMSD metrics for the models including satellite information. $q=0$ for all the models.}\label{tab:MBD_RMS_satellite}
\begin{small}
\begin{tabular*}{\textwidth}{c@{\extracolsep{\fill}}C{0.9cm}C{0.9cm}C{0.9cm}C{0.9cm}C{0.9cm}C{0.9cm}C{0.9cm}C{0.9cm}C{0.9cm}}
\hline
\multicolumn{10}{c}{\textbf{ARMAX-RLS model specification}}\\
\hline
\textbf{window}& medium& medium& medium& medium& medium& small& small& small& large\\
$\mathbf{p}$&  0& 1& 5& 1& 1& 0& 0& 1& 1\\
\textbf{lags}& 0& 0& 0& 1& 5& 0& 5& 1& 1\\
\hline
\textbf{lead time}&\multicolumn{9}{c}{\textbf{relative MBD (\%)}}\\
\cline{1-1}\cline{2-10}
 10~mins& $-0.6$& $-0.3$& $-0.3$& $-0.3$& $-0.3$& $-0.7$& $-0.6$& $-0.3$& $-0.3$\\
 20~mins& $-0.6$& $-0.5$& $-0.5$& $-0.5$& $-0.5$& $-0.7$& $-0.6$& $-0.5$& $-0.4$\\
 30~mins& $-0.7$& $-0.6$& $-0.6$& $-0.6$& $-0.6$& $-0.8$& $-0.7$& $-0.6$& $-0.5$\\
 40~mins& $-0.7$& $-0.6$& $-0.6$& $-0.6$& $-0.6$& $-0.8$& $-0.7$& $-0.7$& $-0.6$\\
 50~mins& $-0.7$& $-0.7$& $-0.7$& $-0.7$& $-0.7$& $-0.8$& $-0.7$& $-0.7$& $-0.7$\\
 60~mins& $-0.8$& $-0.7$& $-0.7$& $-0.7$& $-0.7$& $-0.8$& $-0.8$& $-0.8$& $-0.7$\\
 90~mins& $-0.8$& $-0.8$& $-0.8$& $-0.8$& $-0.8$& $-0.9$& $-0.8$& $-0.9$& $-0.8$\\
120~mins& $-0.8$& $-0.8$& $-0.8$& $-0.8$& $-0.8$& $-0.9$& $-0.8$& $-0.9$& $-0.8$\\
150~mins& $-0.9$& $-0.9$& $-0.8$& $-0.9$& $-0.9$& $-0.9$& $-0.9$& $-1.0$& $-0.8$\\
180~mins& $-0.9$& $-0.9$& $-0.9$& $-0.9$& $-0.9$& $-0.9$& $-0.9$& $-1.0$& $-0.9$\\
220~mins& $-1.0$& $-1.0$& $-1.0$& $-1.0$& $-1.0$& $-1.0$& $-1.0$& $-1.1$& $-1.0$\\
260~mins& $-1.2$& $-1.1$& $-1.1$& $-1.1$& $-1.1$& $-1.2$& $-1.2$& $-1.2$& $-1.1$\\
\hline
\textbf{lead time}&\multicolumn{9}{c}{\textbf{relative RMSD (\%)}}\\
\cline{1-1}\cline{2-10}
 10~mins& $19.0$& $17.5$& $17.6$& $17.5$& $17.5$& $19.7$& $19.1$& $17.7$& $17.7$\\
 20~mins& $19.5$& $19.3$& $19.4$& $19.3$& $19.3$& $20.3$& $19.5$& $19.8$& $19.5$\\
 30~mins& $20.6$& $20.6$& $20.6$& $20.4$& $20.4$& $21.6$& $20.5$& $21.3$& $20.5$\\
 40~mins& $21.8$& $21.9$& $21.9$& $21.7$& $21.7$& $22.9$& $21.7$& $22.7$& $21.5$\\
 50~mins& $23.1$& $23.1$& $23.1$& $23.0$& $23.0$& $24.2$& $23.0$& $24.0$& $22.6$\\
 60~mins& $24.2$& $24.2$& $24.3$& $24.1$& $24.2$& $25.3$& $24.1$& $25.1$& $23.7$\\
 90~mins& $27.1$& $27.2$& $27.2$& $27.1$& $27.1$& $28.1$& $27.1$& $28.0$& $26.7$\\
120~mins& $29.7$& $29.7$& $29.8$& $29.7$& $29.7$& $30.5$& $29.7$& $30.5$& $29.3$\\
150~mins& $31.9$& $32.0$& $32.0$& $32.0$& $32.0$& $32.6$& $32.0$& $32.6$& $31.6$\\
180~mins& $33.9$& $34.0$& $34.0$& $34.0$& $34.0$& $34.5$& $34.0$& $34.5$& $33.6$\\
220~mins& $36.3$& $36.3$& $36.4$& $36.3$& $36.4$& $36.7$& $36.3$& $36.8$& $36.0$\\
260~mins& $38.2$& $38.3$& $38.4$& $38.3$& $38.4$& $38.6$& $38.3$& $38.7$& $38.1$\\
\hline
\end{tabular*}
\end{small}
\end{table*}

\newpage

\begin{table*}[ht!]
\renewcommand{\arraystretch}{1.1}
\centering
\caption{Forecasting skill for the models including satellite information. $q=0$ for all the models.}\label{tab:FS_satellite}
\begin{small}
\begin{tabular*}{\textwidth}{c@{\extracolsep{\fill}}C{0.9cm}C{0.9cm}C{0.9cm}C{0.9cm}C{0.9cm}C{0.9cm}C{0.9cm}C{0.9cm}C{0.9cm}}
\hline
\multicolumn{10}{c}{\textbf{ARMAX-RLS model specification}}\\
\hline
\textbf{window}& medium& medium& medium& medium& medium& small& small& small& large\\
$\mathbf{p}$&  0& 1& 5& 1& 1& 0& 0& 1& 1\\
\textbf{lags}& 0& 0& 0& 1& 5& 0& 5& 1& 1\\
\hline
\textbf{lead time}&\multicolumn{9}{c}{\textbf{FS (\%) using the regular persistence}}\\
\cline{1-1}\cline{2-10}
 10~mins&  $+9.0$& $+16.2$& $+16.1$& $+16.3$& $+16.3$&  $+5.9$& $ +8.8$& $+15.2$& $+15.4$\\
 20~mins& $+21.9$& $+22.6$& $+22.4$& $+22.9$& $+22.9$& $+18.8$& $+21.8$& $+20.8$& $+22.1$\\
 30~mins& $+23.8$& $+23.7$& $+23.6$& $+24.4$& $+24.3$& $+20.0$& $+24.1$& $+21.1$& $+24.2$\\
 40~mins& $+23.3$& $+23.1$& $+23.0$& $+23.7$& $+23.7$& $+19.3$& $+23.6$& $+20.0$& $+24.3$\\
 50~mins& $+22.4$& $+22.3$& $+22.1$& $+22.7$& $+22.7$& $+18.6$& $+22.7$& $+19.2$& $+23.8$\\
 60~mins& $+21.4$& $+21.3$& $+21.2$& $+21.7$& $+21.6$& $+17.9$& $+21.6$& $+18.4$& $+22.9$\\
 90~mins& $+19.5$& $+19.4$& $+19.3$& $+19.6$& $+19.5$& $+16.7$& $+19.5$& $+16.9$& $+20.9$\\
120~mins& $+18.6$& $+18.5$& $+18.4$& $+18.6$& $+18.5$& $+16.4$& $+18.6$& $+16.5$& $+19.7$\\
150~mins& $+18.0$& $+17.9$& $+17.8$& $+17.9$& $+17.8$& $+16.3$& $+17.9$& $+16.3$& $+18.9$\\
180~mins& $+18.2$& $+18.1$& $+18.0$& $+18.1$& $+17.9$& $+16.8$& $+18.0$& $+16.8$& $+18.9$\\
220~mins& $+18.5$& $+18.4$& $+18.2$& $+18.4$& $+18.2$& $+17.5$& $+18.4$& $+17.4$& $+19.0$\\
260~mins& $+19.1$& $+19.0$& $+18.9$& $+19.0$& $+18.8$& $+18.3$& $+18.9$& $+18.2$& $+19.5$\\
\hline
\textbf{lead time}&\multicolumn{9}{c}{\textbf{FS (\%) using the optimal smart persistence}}\\
\cline{1-1}\cline{2-10}
 10~mins&  $+7.4$& $+14.7$& $+14.6$& $+14.8$& $+14.8$&  $+4.2$&  $+7.1$& $+13.7$& $+13.9$\\
 20~mins& $+17.3$& $+18.0$& $+17.8$& $+18.4$& $+18.4$& $+14.0$& $+17.2$& $+16.1$& $+17.5$\\
 30~mins& $+18.7$& $+18.6$& $+18.4$& $+19.3$& $+19.2$& $+14.6$& $+19.0$& $+15.8$& $+19.1$\\
 40~mins& $+18.2$& $+18.0$& $+17.9$& $+18.6$& $+18.6$& $+14.0$& $+18.5$& $+14.7$& $+19.3$\\
 50~mins& $+17.3$& $+17.2$& $+17.0$& $+17.7$& $+17.6$& $+13.3$& $+17.6$& $+13.9$& $+18.8$\\
 60~mins& $+16.6$& $+16.4$& $+16.3$& $+16.8$& $+16.7$& $+12.8$& $+16.8$& $+13.3$& $+18.1$\\
 90~mins& $+15.4$& $+15.2$& $+15.1$& $+15.4$& $+15.3$& $+12.4$& $+15.4$& $+12.6$& $+16.8$\\
120~mins& $+14.7$& $+14.6$& $+14.5$& $+14.7$& $+14.6$& $+12.4$& $+14.7$& $+12.5$& $+15.9$\\
150~mins& $+14.5$& $+14.4$& $+14.3$& $+14.4$& $+14.3$& $+12.7$& $+14.4$& $+12.7$& $+15.4$\\
180~mins& $+14.7$& $+14.6$& $+14.5$& $+14.6$& $+14.5$& $+13.3$& $+14.6$& $+13.3$& $+15.4$\\
220~mins& $+14.9$& $+14.8$& $+14.6$& $+14.7$& $+14.6$& $+13.8$& $+14.7$& $+13.7$& $+15.4$\\
260~mins& $+14.1$& $+14.0$& $+13.9$& $+14.0$& $+13.8$& $+13.3$& $+13.9$& $+13.2$& $+14.5$\\
\hline
\end{tabular*}
\end{small}
\end{table*}

\bibliographystyle{model5-names}
\bibliography{ref_ralonsosuarez.bib}

\end{document}